\documentstyle[12pt]{report}
\def\lsim{\lower0.6ex\vbox{\hbox{$ \buildrel{\textstyle <}\over{\sim}\ $}}}
\def\rsim{\lower0.6ex\vbox{\hbox{$ \buildrel{\textstyle >}\over{\sim}\ $}}}
\def \he3{$^3$He}
\def\he4{$^4$He}

\def\cth{$^{13}$C}
\def\o18{$^{18}$O}
\renewcommand{\baselinestretch}{1.0}

\begin{document}
\noindent
         \hfill OSU-TA-16/95\\
 \vspace{  1.5cm}
\begin{center}
\begin{Large}
\begin{bf}
GALACTIC EVOLUTION OF \\ D AND $^3$He
 INCLUDING \\ STELLAR PRODUCTION OF $^3$He\\
\end{bf} \end{Large}
\end{center}
\begin{center}
\begin{large}
DAVID S. P. DEARBORN$^1$,
 GARY STEIGMAN$^2$\\

AND MONICA TOSI$^3$\\
\end{large}
\end{center}
 \vspace{0.3cm}

 \vspace{6.5cm}
\vfil

\noindent {\footnotesize $^1$Lawrence Livermore National Laboratory L-58,
P. O. Box 808,
Livermore, CA 94550}\\
{\footnotesize $^2$Departments of Physics and Astronomy,
The Ohio State University,
174 West
 18th Avenue, Columbus, OH 43210}\\
{\footnotesize $^3$Osservatorio Astronomico di Bologna, Via Zamboni 33,
40126 Bologna, ITALY}

 \vfil\eject

\centerline{\bf ABSTRACT}

New stellar models which track the production and destruction of
$^3$He (and D) have been evolved for a range of stellar masses
$(0.65\leq M/M_{\odot}\leq 100)$, metallicities $(0.01 \leq Z/Z_{\odot}
\leq 1)$
and initial (main sequence) $^3$He mass fractions $(10^{-5} \leq X_{3,MS}
\leq 10^{-3})$.  Armed with
the $^3$He yields from these stellar models we have followed
the evolution of D and $^3$He using a variety of chemical
evolution models with and without infall of primordial or
processed material.  Production of new $^3$He by the lower
mass stars overwhelms any reasonable primordial contributions
and leads to predicted abundances in the presolar nebula and/or
the present interstellar medium in excess of the observationally
inferred values.  This result, which obtains even for zero primordial D and
$^3$He, and was anticipated by Rood, Steigman
\& Tinsley (1976), is insensitive to the choice of chemical
evolution model; it is driven by the large $^3$He yields from
low mass stars.  In an attempt to ameliorate this problem we
have considered a number of non-standard models in which the
yields from low mass stars have been modified.  Although
several of these non-standard models may be consistent with
the $^3$He data,  they may be inconsistent with observations
of $^{12}$C/$^{13}$C, $^{18}$O and, most seriously, the super-$^3$He
rich planetary nebulae (Rood, Bania \& Wilson 1992).  Even using
the most extreme of these non-standard models (Hogan 1995),
we obtain a generous upper bound to pre-galactic $^3$He:
 X$_{3P} \leq 3.2 \times10^{-5}$  which, nonetheless, leads to a stringent lower
bound to the universal density of nucleons.

 \vfil\eject

\centerline{1. INTRODUCTION}

 To test and constrain models of primordial nucleosynthesis it is necessary
to confront the
predicted abundances with observational data. Such comparisons are
complicated by the
necessity
of extrapolating the abundances inferred ``here and now" (solar system,
ISM, etc.) to their
``there and then" (primordial or pregalactic) universal values. This difficulty
is
somewhat ameliorated
for
$^4$He which is observed in less evolved, low metallicity extragalactic HII
regions (e.g.,
Pagel
et al.\ 1991; Skillman et al.\ 1994; Olive \& Steigman 1995) and for $^7$Li
which is probed in
 metal poor halo stars (Spite \& Spite 1982; Thorburn 1994). In contrast,
$^3$He is only
observed in the solar system (Black 1971, 1972; Geiss \& Reeves 1972; Geiss
1993) and in
Galactic HII regions (Bania, Rood \& Wilson 1987; Balser et al.\ 1994).
Until recently, D,
too, had only been observed ``here and now". A new era in probing deuterium
has dawned with
the identification of possible absorption of QSO light by high redshift D
(Carswell et al.\ 1994; Songaila et al.\ 1994;
Tytler 1995). However, ``wrong velocity" hydrogen absorbers may
masquerade as deuterium (Carswell et al.\ 1994; Songaila et al.\ 1994;
Steigman 1994) so that
much more data are required before such observations may be used to fix the
nearly primordial
abundance of D.  Thus, to infer the pregalactic abundances of D and $^3$He
it is currently necessary to have recourse to models of galactic chemical
 evolution. Unfortunately, the
uncertainties and/or lack of uniqueness of such models compound
 the
observational uncertainties.

The evolution of D is straightforward since when incorporated in a star it is
burned (to
$^3$He) during the pre-main sequence evolution. If the ``virgin" fraction
of the ISM (either
today or at the time of formation of the solar system 4.5 Gyr ago) were
known, the
primordial D abundance could be inferred from ISM or solar system
observations. A very large
class of chemical evolution models (Audouze \& Tinsley 1974; Tosi 1988 a,b;
Vangioni-Flam \&
Audouze 1988; Matteucci \& Fran\c cois 1989; Steigman \& Tosi 1992),
constrained by heavy
element abundances, abundance gradients, abundance ratios (primary vs.\
secondary) and
cosmochronometers, find that $\sim 1/3 - 2/3$ of the ISM (now, $t_0$ or, at
the time of the
solar system formation, $t_\odot$) has never been through stars. However,
``designer
models" with larger D destruction (Vangioni-Flam, Olive \& Prantzos 1994;
Olive et al.\ 1995;
Scully \& Olive 1995) do exist (but see, e.g., Edmunds 1994 and Prantzos 1995
for their consistency problems).

The galactic evolution of $^3$He is complex. Any prestellar $^3$He is
enhanced by the
pre-main sequence burning of prestellar D. Thus, when a star reaches the
main sequence the
$^3$He mass fraction exceeds that in the prestellar nebula [$X_{3,MS} = (X_3 + 3
X_2/2)_{pre}$; in what follows we write $X_i$ for mass fractions and $y_i$
for ratios by
number to hydrogen; $y_{ij} \equiv y_i + y_j]$. $^3$He survives nuclear
burning in the
cooler outer layers of stars, is destroyed in the hotter interiors and,
especially in low
mass stars, is synthesized via hydrogen burning. Thus, depending on the
mass (and, to a
lesser extent the metallicity) of a star, $^3$He is preserved, produced or
destroyed. To
understand the galactic evolution of $^3$He, it is necessary to first
understand its stellar
evolution (Iben 1967; Rood 1972; Dearborn, Schramm \& Steigman 1986;
Vassiliadis \& Wood 1993).
However, since {\it some} $^3$He survives stellar
processing and, prestellar D is burned to $^3$He, the evolution of $^3$He
(or, D + $^3$He)
is less rapid than that of D alone (Yang et al.\ 1984; hereafter YTSSO).
Thus, observations
of D and $^3$He (e.g., in the solar system) may be used to infer an {\it
upper bound} to
primordial D and/or D + $^3$He (YTSSO; Walker et al.\ 1991; hereafter
WSSOK). As a result,
models which destroy D more efficiently are to some extent constrained  by
the requirement
that solar system and/or ISM $^3$He not be overproduced.

Rood, Steigman \& Tinsley (1976; hereafter RST) first included stellar
production of $^3$He
in numerical models of galactic evolution. RST found that stellar
production led to a rapid
increase in $X_3$ from $t_{\odot} \ {\rm to}\ t_0$ which seemed in conflict
with the then
current observations (only bounds) and concluded that it seemed unlikely
that $^3$He could
be used cosmologically. As we shall see, this problem persists today (Olive et
al. 1995).

In a previous study (Steigman \& Tosi 1992; hereafter ST92) we used the $^3$He
survival from Dearborn,
Schramm \& Steigman (1986) in models of galactic evolution consistent
with a large
number of observational constraints (Tosi 1988a) to track the evolution of D
and $^3$He.
Not including $^3$He production in low mass stars, we found that $X_{3P}
\approx X_{3\odot} \approx X_{30}$ for a wide range of choices of initial
(primordial)
abundances for D and $^3$He. Similar results have been found, e.g., by
Fields (1995). At
the same time, the $^3$He HII region data (Bania, Rood \& Wilson 1987;
Rood, Bania \& Wilson
1992) have been supplemented by new and important observations of
super--$^3$He rich material in a few planetary nebulae (Balser et al.\ 1994,
Rood et al.\ 1995, hereinafter RBWB95).
Since the gas in PNe reflects the chemical
composition of the ejected outer layers of their central stars, where the
original $^3$He abundance has certainly been modified by stellar processing,
their large $^3$He abundances should not be taken as representative of the ISM
abundances
at the time of the formation of their central stars but, of the
effect of stellar nucleosynthesis in low mass stars. The observed abundances
lend support to the estimates of
newly synthesized $^3$He $(X_{3*} \approx 0.7 - 7 \times 10^{-3})$ in low
mass stars $(0.6 -
2.3 M_{\odot})$ in the pioneering work of Iben (1967) and Rood (1972) and
makes timely a
reinvestigation of the galactic evolution of $^3$He (along with D) {\it
including} stellar
production of $^3$He (Galli et al.\ 1995, Tosi, Steigman \& Dearborn 1995,
Olive et al.\ 1995).
To this aim, we have computed an extensive grid of stellar evolution models
with varying initial abundances
of $^3$He and of the overall metallicity Z, and we have reexamined
the galactic evolution of D and $^3He$ in the framework of such stellar
models.

In \S 2 we describe the stellar models we've evolved to follow the
evolution of $^3$He and
present our results for the survival/production of $^3$He as a function of
stellar mass
$(0.65 \le M/M_{\odot} \le 100)$, metallicity $(Z = 0.02, \ 0.002,\
0.0002)$ and initial
(``main sequence" $\Rightarrow $D + $^3$He) $^3$He\ $(10^5 X_{3,MS} = 1.0,\
5.0,\ 10,\ 21,\
100)$.
 In \S 3 and \S 4 we use the stellar results in a series of chemical evolution
models (Tosi 1988a,b;
ST92) to follow the galactic evolution of D and $^3$He. In \S 5 we further
follow the $^3$He evolution modifying the stellar yields.  Sections 6 \& 7 are
reserved for discussions and conclusions.

\vspace {1.5cm}
\centerline{2. STELLAR EVOLUTION OF $^3$He}

The evolution of stars with $M = 0.65 M_{\odot} \ {\rm to}\ M = 100
M_{\odot}$ and three
metalicities (``Pop I": $X =  0.70,\ Z = 0.02,$ ``Pop I 1/2": $X= 0.76,\ Z
= 0.002,$ ``Pop
II": $X = 0.76,\ Z = 0.0002)$ was followed using the code described in
Dearborn, Griest \&
Raffelt (1991) which is derived from that of Eggelton (1967, 1968). OPAL
opacities
(Iglesias \& Rogers 1990, 1992) are used for temperatures above 6000K and
Los Alamos
opacities (Heubner et al. 1977), including the contribution of molecules, for
lower
temperatures. Our code follows $^3$He production and destruction in all
hydrogen burning
regions and a post-processor is used to calculate the nucleosynthesis of D
and the isotopes
of Li, Be, C, N, O and several other elements. Calculations  with this
post-processor
confirmed the pre-main sequence conversion of D to $^3$He. The mixing
length is chosen to
give the correct solar radius and this model predicts a neutrino flux (for
the chlorine
experiment) of 7.9 SNU and the p-mode spectrum matches that of Bahcall \&
Ulrich (1988).
For the Pop I 1/2 and Pop II models, a slightly larger mixing length is
required to fit the
color-magnitude data for M30 (Bolte 1994). However, $^3$He evolution is
insensitive to any
reasonable choice of mixing length.

All models began in the pre-main sequence on the Hayashi track. The final
evolutionary state
depended on the mass of the star. Stars with $0.65 \le M/M_{\odot} \le
2.00$ were evolved up
to the helium flash. For one model $(1.5 M_{\odot})$ the evolution was
followed through the
helium flash and up the asymptotic giant branch (AGB). This additional
evolution (without thermal pulses) changed the
$^3$He mass fraction in the envelope by less than 1\%, demonstrating that
in the absence of
thermal pulses, the envelope abundance of $^3$He is insensitive to the
subsequent (post
helium flash) evolution of these stars.
Intermediate mass stars $(2.5 \le
M/M_{\odot} \le
8.0)$ were all evolved to the AGB, to a point where mass loss (and possibly
Planetary Nebula
formation) dominates their remaining evolution. All massive star models ($M
\ge 10
M_{\odot}$) were evolved into carbon burning and are within a few thousand
years of core
collapse, leaving insufficient time for significant changes in the
composition of their
hydrogen envelopes. We have not allowed for winds and this suggests that,
for the more
massive stars, we have underestimated the $^3$He returned to the
ISM (see, e.g., Dearborn, Schramm \& Steigman 1986).

Since D is completely burned to $^3$He in the pre-main sequence, our ``main
sequence" (MS)
$^3$He abundance reflects the prestellar contribution from D and $^3$He:
$X_{3,MS} = X_3 + 3 X_2/2$.
To explore the sensitivity of our $^3$He yields (envelope fractions
$X_3 (env))$ to
$X_{3,MS}$ we have chosen five values of $X_{3,MS} \ {\rm from}\  1 \times
10 ^{-5} \ {\rm
to}\ 1
\times 10 ^{-3}$. The top panel of Figure 1 shows our Pop I yields $X_{3f}
\equiv X_3(env)$ as a
function of the initial stellar mass for different choices of $X_{3,MS}$.
For lower mass
stars production dominates and, when convection on the giant branch
homogenizes the
envelope, the $^3$He available to be returned to the ISM is strongly
enhanced. Without
substantial mass loss, these lower mass stars remain near the Hayashi track
during helium
burning and beyond. In the absence of a hot bottom convection zone
stimulated by thermal
pulses, there is no destruction of this ``new" $^3$He. For more massive
stars the lifetimes
are shorter resulting in less $^3$He production in the temperature regimes
where this
occurs. Note that if these stars started with no initial $^3$He they would,
in fact, be net producers of $^3$He;
in this case $g_3 \equiv X_{3f}/X_{3,MS}$ would
diverge. However, for
the range of $X_{3,MS}$ of interest $(10^{-5} - 10^{-3})$, the small
production of $^3$He in
more massive stars does not compensate for the destruction of initial $^3$He.

A striking feature of the 15 and 25$M_{\odot}$ Pop I models is the dip in
$X_{3f}$. The high
envelope opacity allowed these models to homogenize the envelope on the
giant branch prior
to helium burning. The blue loop that occurred during helium burning then
led to a second
epoch of $^3$He destruction. In contrast, the 50 and 100$M_{\odot}$ models
completed their
helium burning before evolving to the red thus leading to  the ``recovery"
of $X_{3f}$ seen in the top panel of Figure 1. Similarly, the lower
metallicity (Z=0.002 and Z=0.0002) 25$M_{\odot}$ models
did not evolve all the way to the red giant branch prior to helium burning
and, therefore,
these stars only experienced $^3$He destruction on the main sequence (see
bottom panel of Fig.1).
For Pop I 1/2 and Pop II stars the lower opacity results in a more compact,
warmer structure
(similar to that of a slightly more massive Pop I star). At a fixed mass
this leads to a
shorter lifetime resulting in less production and more destruction of
$^3$He. However, the
lower metallicity 15 \& 25$M_{\odot}$ models do not experience a second
epoch of $^3$He
destruction. As a result, they show less $^3$He destruction than the Pop I
models (i.e., no
dip in $X_{3f} \ {\rm vs.}\ X_{3,MS})$. By comparing the two panels of Fig.1,
it is apparent that the amount of $^3$He ejected by stars depends
much more on its main sequence $^3$He abundance than on the overall stellar
metallicity.

In Figure 2 we display our Z=0.02 (Pop I) results from a different perspective
by
showing the final
yield ($X_{3f}$) as a function of the initial abundance $(X_{3,MS})$ for $
1 \le M/M_{\odot}
\le 100$. For $M = 1 M_{\odot}$, production dominates and $X_{3f}$ is
(nearly) independent of
$X_{3,MS}$ (except at the highest $X_{3,MS})$. In contrast, for $M = 100 M
_{\odot}$
production is negligible and $X_{3f}$ varies linearly with $X_{3,MS}$. The
transition from
low to high mass models is seen for $M$ = 3 and $8M_{\odot}$.
For $0.65 \le M/M_{\odot} \le 2.5$, where production dominates, we find
$X_{3f}
\approx (M_{\odot} / M)^{2.2} \times 10 ^{-3}$. For
more massive stars the
relation between
$X_{3f} \ {\rm and}\ X_{3,MS}$ is more complex. As $X_{3,MS}$ increases,
$^3$He production is
relatively less important resulting in the curvature seen in Figure 2 for $M
=$ 3 and  8
$M_{\odot}$. For high masses where production is negligible, $X_{3f} \approx
0.33 X_{3,MS}$.

In the absence of $^3$He production it is interesting to consider the
$^3$He ``survival
fraction" $g_3 \equiv X_{3f}/X_{3,MS}$ (Dearborn, Schramm \& Steigman
1986; YTSSO; ST92). In
the presence of $^3$He production $g_3$ is less useful and, potentially
misleading since the
yield is not proportional to $X_{3,MS}$. Nonetheless, for comparison with
previously
published results, $g_3$ vs.\ M for our five choices of $X_{3,MS}$ is shown
in Figure 3. The
divergence of the curves at low M reflects the increasing importance of
$^3$He production which  sets in first (at the largest M)
for the lowest values of $X_{3,MS}$.
Note that in
the limit that $X_{3,MS}$ goes to zero, $g_3$ diverges. In Figure 4 is shown
$\langle g_3
\rangle$, the value of $g_3$ averaged over Tinsley's (1980) IMF as a
function of the lower mass limit m$_l$ of the IMF.
For gas incorporated in a generation of stars, as time
increases
material is returned from lower mass stars and so Fig.\ 4 provides a
picture of the time-evolution
of the $^3$He ``survival".  Note that for all $m_l$, $\langle g_3
\rangle \ \rsim \ 0.3$. Depending on the initial abundance,
production dominates, $\langle g_3 \rangle \ \rsim 1$, for $m_l$ up to
7$M_{\odot}$. In terms of galactic evolution this implies that the ISM starts
to be enriched in $^3$He as soon as stars less massive than $3-7 M_{\odot}$
(depending on the initial $^3$He abundance) start to die. We note
again that stars above $\sim 25 M_{\odot}$ are likely to experience mass
loss which will
return unburnt $^3$He to the ISM. For massive stars our neglect of mass
loss results
in an underestimate of $g_3$.

The trends in our results are easy to understand. During the hydrogen
burning phase $^3$He
is produced in the core of stars of all masses. However, with time, the
core temperature
(which depends on stellar mass) increases to values where $^3$He is burned.
Thus, new $^3$He
survives only in a radiative shell adjacent to the convective core and,
from this shell, may
later be dredged up to the surface where it will mix with the prestellar
$(MS)\ ^3$He. The
competition between net production or destruction then depends not only on
$X_{3,MS}$ but,
on the size of the shell and the amount of new $^3$He there, and on the
depth of the
dredge-ups (i.e., do they bring to the surface material which is enhanced
or depleted in
$^3$He). In low mass stars much  new $^3$He is produced and most of it
survives to
dominate over (reasonably small values of) $X_{3,MS}$. In contrast, for
massive stars little
new $^3$He is synthesized in the core, destruction is efficient and the
$^3$He preservation
shell is small so that the initial $^3$He dominates.

We note that  the Eggleton type code we use permits larger time steps on the
AGB which tends
to suppress thermal pulses. However, when short timesteps are enforced,
these pulses occur.
Such pulses are plausible sites for the lithium enhancements -- produced at
the expense of
 $^3$He -- observed in some S-type stars. Although it doesn't require much
$^3$He depletion
to yield huge lithium enhancements, since the lithium is fragile it is
possible that the
lithium enhancement attains a steady state with $^3$He processed to
$^4$He. A limit on
such $^3$He processing might follow from the fact that a very fragile
nucleus, $^{18}$O,
survives the dredge up in normal carbon stars (Dearborn 1992). Since $^3$He
is destroyed at
higher temperatures than those for $^{18}$O, the lack of $^{18}$O depletion
suggests that
$^3$He is not strongly depleted in normal carbon stars.

A similar argument applies to the main sequence mixing proposed to explain
the low
$^{12}$C/\cth \ ratios observed in red giants (Dearborn \& Eggleton 1976;
Dearborn 1992). In
red giants  \o18 appears to be independent of the \cth \ enhancement,
perhaps due to
stabilization by a molecular weight gradient. Since it is difficult to mix
with a region of
depleted $^3$He without modifying {\o18}, it is suggestive that $^3$He is
not destroyed in
these stars.  In contrast, J-type carbon stars have CNO-equilibrium values
for $^{12}$C/\cth
\ and do show \o18 depletions. In such stars $^3$He is likely destroyed
throughout the
envelope. If most carbon stars pass through such a stage (rather than these
representing a
separate class of stars), little $^3$He may survive. Pinsonneault (Private
Communication) has
noted that for solar metallicity stars, the open cluster M67 seems to be
the dividing line
between stars which do not undergo giant branch mixing and those which do
suggesting that
mixing for stars with $M\ \rsim 1.3 - 1.5 M_{\odot}$ is inconsistent with
the $^{12}$C/\cth \ data.

  It is interesting to compare our results with the recent work of Vassiliadis
\& Wood (1993).  We both agree that in the low mass stars there is no
significant
change in $^3$He between the first and second dredge up.  Further,
Vassiliadis \&
Wood calculate thermal pulses and, for stars below 5 solar masses, they find
no change in the $^3$He abundance.  Our mixing length approximations are
nearly the same and they included a wind which stripped the envelope on the
AGB while we did not.   Quantitatively, our yields and theirs are in
excellent agreement (provided that their $\theta (^3$He) is a mass
fraction) despite
the fact that their opacities are different (higher) than ours and they used
slightly older nuclear reaction rates.

 \vspace{1.5cm}

\centerline{3. THE GALACTIC EVOLUTION OF D}

As in ST92 we have followed the evolution of D and $^3$He for the two
``best" chemical evolution models for the galactic disk identified in Tosi
(1988a) along with a third model in
which, for comparison, infall is absent. Model 1, the ``best" model (Tosi
1988a; Giovagnoli
\& Tosi 1995), consistent with the major
 observational constraints, has  an exponentially decreasing SFR (with a 15
Gyr e-folding time),  depending on both the gas and total mass density
currently observed in each ring,
a constant (in time), uniform (in space) infall rate of 0.004
$M_{\odot}/kpc^2/yr$ and uses
Tinsley's (1980) IMF. Model 25, the ``second best" model (Tosi 1988a) also has
an
exponentially decreasing SFR with, however, a 5 Gyr e-folding time, an
effectively
constant (e-folding time of 100 Gyr) infall rate of uniform density 0.002
$M_{\odot}/kpc^2/yr $ and Tinsley's (1980) IMF. Our comparison No Infall
(NI) model uses
the same IMF and SFR e-folding time as Model 1 but is normalized so as to
reproduce the
current SFR and gas/total mass distributions with galactocentric distance.
We recall, however, that such a model does not reproduce the major features
(distribution with time and galactocentric distance of the chemical abundances)
observed in the galactic disk.

Normally, our
models adopt 13 Gyr for the present epoch and 8.5 Gyr for the formation of
the solar system.
To explore the sensitivity to the age of the model, we have run some models
(indicated by a
subscript 10) where the present epoch is 10 Gyr and solar system formation
is at 5.5 Gyr.

Since infall plays an important role in Models 1 \& 25, it is necessary to
specify the
chemical composition of the infalling gas. We have considered models with
primordial
infall ($Z_{inf} = 0$) for which $X_{2inf} = X_{2P}$ and $X_{3inf} =
X_{3P}$ as well as
models with partially processed infalling material. In the latter cases we
have adopted  $Z_{inf} = 0.2 Z_{\odot}$, a value low enough to preserve the
infall dilution efficiency and keep the model predictions in agreement with
the observational constraints in the solar neighbourhood and in the whole
disk (Tosi 1988b, Matteucci \& Fran{\c c}ois 1989). When $Z_{inf}= 0.2
Z_{\odot}$, the infall abundance of D is certainly lower than primordial due
to stellar processing, whereas the $^3$He abundance in principle can be either
lower or higher depending on the mass of the stars contributing to the
enrichment of the accreted gas. We have thus considered models
 with $X_{2inf} = 0.7 X_{2P}$ or
$X_{2inf} = 0.8 X_{2P}$ and $X_{3inf} = 0.8 X_{3P}$ or $X_{3inf} = X_{3P}$ or
$X_{3inf} = 1.2 X_{3P}$.
>From a series of model checks we find that
larger deviations of the infall abundances of D
from the primordial value would be rather improbable with an
overall metallicity $Z_{inf} = 0.2 Z_{\odot}$.

To explore the sensitivity of our results to the
primordial values we have run each model for a range of choices of $X_{2P}$
and $X_{3P} \ (0
\le 10^5 X_{2P} \le 9.0;\ 0 \le 10^5 X_{3P} \le 4.0$).
In Table 1 we list the set of models explored.

The evolution of deuterium is straightforward since any D incorporated in
stars is
destroyed. Thus, the D survival factor, $f_2 = X_2/X_{2P}$, is identical to the
fraction of
the ISM that has never been through stars. $f_2$, then, is independent of
$X_{2P}$ and purely
reflects the chemical evolution model. In Figure 5 we show $f_2$ evaluated
at the solar ring
$(R=8kpc)$ as a function of time. For each model (1 \& 25) we show the
differences between
primordial infall $(Z_{inf} = 0,\ X_{2inf} = X_{2P})$ and non-primordial
infall $(Z_{inf} =
0.2 Z_{\odot},\ X_{2P} = 0.8 X_{2P})$ as well as the corresponding
no-infall (NI) model.
We also compare in Figure 5 models whose present age is 13 Gyr with 10 Gyr
models. The
models with non-primordial infall have less D refueling of the ISM and show
a steeper decrease with time of $f_2$ than models with primordial infall.
The NI model destroys
D more slowly at
first but, without replenishment of ISM D via infall, eventually has the
largest D
depletion. The short lifetime models (10 Gyr) have less time to destroy D
but, to reproduce
present observations have higher initial mass and SFR than the 13 Gyr
models. The net result
of this balancing act is that the 10 Gyr (infall) models have higher $f_{2
\odot}$ but lower
$f_{20}$; the NI models have larger $f_2$ correlating with lower age. For
Models 1 \& 25
the entire ranges are: $0.49 \le f_{2 \odot} \le 0.73;\ 0.43 \le f_{20}
\le 0.62$; for the
NI models: $0.70 \le f_{2 \odot} \le 0.78;\ 0.30 \le f_{20} \le 0.36$.
Thus, for our
range of models D astration is modest, typically by a factor of 1.3 to 2.0
at $t_{\odot}$ and
a factor of 1.6 to 3.3 at
$t_{0}$. Our Models 1 \& 25 have very little evolution in $f_2$ from
$t_{\odot}\ {\rm to}\
t_{0}\ (0.9 \le f_{20}/f_{2\odot} \le 1.2 \ {\rm for}\ t_{0} = 13\
{\rm Gyr};\ 1.3 \le
f_{20}/f_{2 \odot} \le 1.5 \ {\rm for}\ t_{0} = 10\ $Gyr). This is
entirely consistent
with the solar system and ISM data: $X_{20}/X_{2\odot} = 1.6 \pm 0.6$
(e.g., Steigman \&
Tosi 1995).

In Figure 6 we show the time evolution of $X_2$ in the solar ring
for Models 1, 25 \& NI for
$t_{0} = 13,\ 10\
$Gyr. Also shown in Figure 6 are the $2 \sigma$
ranges of the solar system (Geiss 1993; Steigman \& Tosi 1995)
and ISM (Linsky et
al.\ 1992; Steigman \& Tosi 1995) D abundances. For Models 1 \& 25 the ISM
constraint
dominates, limiting primordial D to $X_{2P} \le 6 \times 10^{-5} \ {\rm
(for}\ Y_P\ \lsim 0.25, \ y_{2P}\ \lsim 4 \times 10^{-5})$.
Even for the NI model we find a restrictive bound $X_{2P} \le 9 \times
10^{-5} \ (y_{2P}\
\lsim 6 \times 10^ {-5})$.

In Figure 7 is shown the predicted radial distribution of D/H at present.
The positive
gradient is a natural consequence of larger D destruction in regions with
larger SFR. The
``data" are from Wannier (1980) {\it but} have been {\it divided by a
factor of 100} to
facilitate comparison; while they appear to reflect the expected radial
distribution, they
are some two orders of magnitude {\it too large}, possibly due to the chemical
fractionation of deuterated molecules.

With the adoption of a chemical evolution model (e.g., 1, 25 or NI; $t_0 =
13 $ or 10 Gyr;
$X_{2inf}/X_{2P} = 1.0 \ {\rm or}\ 0.8$) the evolution of deuterium (with
time and location
in the Galaxy) is completely determined. Comparison with solar system and
ISM observations
then fixes (bounds) the primordial D abundance. For our ``normal" models
(not the NI models)
we find $2.7 \le 10^5 X_{2P} \le 6.6 \ ({\rm for }\ X_P\ \rsim 0.75, \ 1.8\
\lsim 10^5 y_{2P} \
\lsim 4.4)$ which, for standard big bang nucleosynthesis (Walker et al.\
1991; Thomas et al.\
1995) bounds the nucleon-to-photon ratio $\eta: \ 4.1 \ \lsim \eta_{10} \
\lsim 7.1 \ (\eta_{10}  = 10^{10} \eta = 10^{10} n_N / n_{\gamma})$. Our
(artificial) NI models prefer
 somewhat higher primordial D: $4.7 \le 10^5X_{2P} \le 8.8 \ (3.2 \ \lsim 10^5
y_{2P} \ \lsim 5.9;
\ 3.5 \ \lsim \ \eta_{10} \ \lsim \ 5.1)$. The modest astration we find
is similar to many previous results (Audouze \& Tinsley 1974; ST92;
Galli et al. 1995; Fields 1995).
Nonetheless, it may be possible to construct chemical evolution
models with more destruction of D although consistency with all the
observational constraints is quite improbable
(Vangioni-Flam \& Audouze 1988; Vangioni-Flam, Olive \& Prantzos 1994;
Olive et al.\
1995; Olive \& Scully 1995). An important test of all models is the
evolution of $^3$He (RST).

 \vspace{1.5cm}
\centerline {4. THE GALACTIC EVOLUTION OF $^3$He: STANDARD MODELS}

  The evolution of $^3$He is much more complex than that of D since $^3$He
may be destroyed, preserved and produced in differing proportions in stars of
differing masses.  Here, we have used the stellar results described in \S 2
in concert with the large number of models summarized in Table 1.  The key
difference with our earlier work (ST92), which utilized similar models, is
our allowance here for the production of new $^3$He synthesized in lower mass
stars.  Following the results of \S 2,
in running the models, at each time step the adopted stellar yields are those
corresponding to the $X_{3f}$ of the dying stars.
As anticipated by RST, the effect of $^3$He production is large,
dominating the evolution of $^3$He.

In Figure 8 the evolution of $^3$He  is shown for a wide
variety of choices of $X_{2P}, X_{3P}, X_{2inf}$, $X_{3inf}$ for Model 1
and a present age of 13 Gyr. The vertical bar at 8.5 Gyr corresponds to the
2$\sigma$ range of abundances derived in the solar system (Geiss 1993). Given
the lack of $^3$He determinations for the local ISM, the vertical bar at 13
Gyr corresponds to the 2$\sigma$ range of abundances derived by RBWB95 for
HII regions between  6.4 and 10.3 kpc from the galactic center (but excluding
W3 which definitely lies outside the average distribution and is the worst
case for pressure broadening corrections).
In the upper panel of Figure 8, the problem of
excessive $^3$He production identified by RST is clear. Using the stellar yields
from \S 2, it may be seen that there are {\bf no} choices of $X_{2P}$ and
$X_{3P}$ consistent with the range of $^3$He abundances inferred from the solar
system and/or ISM data.  After the first few Gyr of evolution the contribution
from newly synthesized $^3$He overwhelms any primordial D+$^3$He and, even when
$X_{2P} = X_{3P}$ = 0, the $^3$He abundances after 8.5 (13) Gyr are in excess of
the solar system and ISM upper bounds.

  The trends displayed in the upper panel of Figure 8 are easy to understand.
For fixed choices of $X_{3P}$, $X_{3}$(t) increases with increasing $X_{2P}$
since any prestellar D is burned to $^3$He, enhancing the main sequence
abundance of $^3$He.  Models which begin with higher $X_{3P}$ always have
higher $^3$He
abundances although, with time, the differences are reduced by the emerging
dominance of the newly synthesized $^3$He.

  The upper four curves in the bottom panel of Figure 8 demonstrate the effect
of the different choices for the infall abundances.  However, these differences
are small compared to the overall enhancement by stellar produced $^3$He.  As
in the upper panel, even in the absence of any primordial D or $^3$He, the
predicted $^3$He abundances at the time or formation of the solar system and/or
at present are in excess of the observational upper bounds.

  Thus, our results confirm - with a vengeance - the RST identified problem of
$^3$He overproduction when the contribution of new $^3$He from low mass stars is
included in models of galactic chemical evolution (see also Galli et al.
1995).  This conclusion is not
modified when we used Model 25, the NI model or, for any of these models with
a 10 Gyr disk age.   We
have, therefore, considered how we might have to modify the contribution of
new $^3$He from the low mass stars in order to reconcile $^3$He evolution
with the observational data.

 \vspace{1.5cm}
\centerline {5. THE GALACTIC EVOLUTION OF $^3$He: NON-STANDARD MODELS}

  As outlined in Table 1, in addition to our standard models (with the yields
from \S 2), we have considered several non-standard models by
modifying the $^3$He yields from low mass stars.  As may be seen from the
bottom
panel of Figure
8 and in both panels of Figure 9, there are a variety of possible solutions
to the problem of $^3$He overproduction.
The rapid increase in $^3$He is most curtailed in models VI which follow the
suggestion of Hogan (1995) that in stars less than 2.5 M$_{\odot}$, $^3$He is
destroyed on the giant branch before it can be returned to the ISM and the
final envelopes of these stars therefore only contain an abundance $X_{3f}$
corresponding to
the equilibrium value $^3$He/H=1$\times 10^{-5}$.
However, Hogan's (1995) suggestion is in conflict with the normal
$^{12}$C/\cth \
ratios observed in stars more massive than 1.3-1.5 M$_{\odot}$ (Pinsonneault,
Private Communication), as well as the observations of lithium (more fragile
than $^3$He) in some of them.  Even worse, the observations of excess $^3$He
(X$_3$ of order 10$^{-3}$) in three planetary nebulae (Rood, Bania \& Wilson
1992; RBWB95) appear to confirm that stars around
1.5 M$_{\odot}$ are efficient $^3$He
producers, in conflict with Hogan's proposal. Charbonnel (1995) has recently
argued that the deep convective mixing responsible for the $^3$He destruction
in Hogan's suggestion takes place in stars experiencing the helium flash and
gives an upper mass limit of 2 M$_{\odot}$ to the phenomenon.  This would
reconcile the deep mixing $^3$He destruction with the high PNe abundances if
the initial mass of the PNe progenitor was at least 2 M$_{\odot}$.  However, 
the three well studied PNe
have presumably originated from stars of initial mass 1.5 M$_{\odot}$
(Stanghellini, 1995 private communication), thus lying in the range of stellar
masses which should deplete and not enhance $^3$He in the deep mixing
hypothesis.

  To try to retain some of the benefit manifest in models VI, we have allowed
$^3$He to be reduced to its equilibrium value in lower
mass stars ($\leq$ 1 M$_{\odot}$ in models V, as suggested by Pinsonneault,
1995, private communication; and $\leq$ 1.6 M$_{\odot}$ in models VII).
However, since such low mass stars have long lifetimes, the reduction in
$^3$He is only effective during recent epochs; solar system $^3$He is still
overproduced in both cases V and VII (see bottom panel of Fig.8 where the two
curves are so close that only one is shown).  Even assuming a total
$^3$He destruction in stars below
1 M$_{\odot}$ (i.e. $X_{3f}$=0, case IV) does not solve this inconsistency,
since it provides results indistinguishable from case V.

 Any other case, with upper mass cutoff for Hogan's $^3$He
destruction intermediate between 1 and 2.5 would either be inconsistent
with the large abundance observed in PNe (if the cutoff is larger than
1.6 M$_{\odot}$) or inconsistent with the ``low" solar system abundances (if
it is smaller than 2 M$_{\odot}$).
We have therefore tested some models assuming that $^3$He is reduced
to its equilibrium value in stars with mass between an arbitrary lower cutoff
and 2.5 M$_{\odot}$. If the lower mass cutoff is around 1.3 M$_{\odot}$ (case
VIII), smaller mass stars still produce large $^3$He consistent with that
observed in PNe and
the ISM abundances predicted at the various epochs are consistent with
the corresponding observed values (see Figs 8 and 9).

  In the models labelled II and III we have - arbitrarily - ignored new $^3$He
production and set g$_{3}$ = 1
for M below 2 M$_{\odot}$ (case II) or, between 1 and 2 M$_{\odot}$
(case III; see, e.g., Wasserburg, Boothroyd \& Sackmann 1995).  As may be
seen in Figure 8, models III are consistent with solar system and ISM data
provided that the primordial abundances of D and/or $^3$He are not too large.

  Although some of the non-standard models II-VIII may avoid overproduction
of
presolar and/or ISM $^3$He the spatial distribution of $^3$He observed in
galactic HII regions (Bania, Rood \& Wilson 1987; Rood, Bania \& Wilson 1992;
Balser et al. 1994; RBWB95) provides an important constraint on all
such models.  The data are puzzling (see, e.g., Olive et al. 1995), exhibiting
no very well defined trend of $^3$He/H with galactocentric distance R (see
Figure 10).  Indeed, in contrast to the theoretical expectation that $^3$He/H
should decrease with R (where the SFR is highest - in the inner galaxy -
$^3$He/H
should also be highest), the data hint at the opposite trend (see Figure 10)
typical of elements like H and D which are destroyed and not produced by
stellar nucleosythesis.
The only models which avoid a decreasing $^3$He vs. R relation are those with
case VI yields, but cases II and, perhaps, III and VIII may also be
consistent with the observed radial distribution, once all the
uncertainties are taken into account.

We note that, despite the apparent conflicts of Hogan's (1995) suggestion
with the $^{12}$C/\cth, lithium and
planetary nebulae data, comparison between case VI and the observational data
still leads to a significant upper bound to primordial $^3$He.
For case VI we find that X$_{3\odot}/X_{3P}$ = 1.4
so that for X$_{3\odot} \leq 4.5 \times 10^{-5}$ (Geiss 1993; ST95), X$_{3P}
\leq 3.2 \times 10^{-5}$ (y$_{3P} \leq 1.4 \times 10^{-5}$).

 \vspace{1.8cm}
\centerline {6. DISCUSSION}

  The evolution of deuterium depends solely on the chemical evolution model, and
the D survival fraction, $f_{2}$(t), is independent of the primordial abundance
and of any stellar uncertainties since D is fully destroyed during the
pre-main sequence evolution.
For our ``best" models (1 \& 25; Tosi 1988a,b), D is
destroyed by only a modest factor (1.4 - 2.0) by the time of formation of the
solar system and, by a slightly larger factor (1.6 - 2.4) up to the present
epoch.  Although the artifical NI models permit somewhat more destruction
(by a factor of 2.8 - 3.3) by the present epoch, they actually destroy less D
(1.3 - 1.4) up to the time of the formation of the solar system.  The solar
system and ISM data (Geiss 1993; Linsky et al. 1992) may be used
along with models 1 \& 25 (for a present age of either 10 or 13 Gyr) to bound
(at the 95\% CL) the primordial D abundance: 2.7 $\leq 10^5$X$_{2P}\leq$ 6.6
(1.8 $\leq 10^5$y$_{2P} \leq$ 4.4).  Comparing to the predictions of BBN (e.g.,
WSSOK) permits us to bound
the universal ratio of nucleons to photons: 4.1 $\leq
\eta_{10} \leq$ 7.1.  For our more extreme NI models we find: 4.7 $\leq 10^5
$X$_{2P} \leq$ 8.8 (3.2 $\leq 10^5$y$_{2P} \leq$ 5.9) and 3.5 $\leq \eta_{10}
\leq$ 5.1.

  In contrast to D, the evolution of $^3$He is much more sensitive to the
details of the chemical evolution model and, especially, to the physics of
stellar structure and evolution.  A key component in our analysis here has been
the computation of a new and extensive grid of stellar models covering a wide
range of masses (0.65 $\leq$ M/M$_{\odot} \leq$ 100), heavy element abundances
(0.01 $\leq$ Z/Z$_{\odot} \leq$ 1) and initial (main sequence) $^3$He abundances
($10^{-5} \leq$ X$_{3,MS} \leq 10^{-3}$).  We have then followed the $^3$He
evolution for models 1, 25 \& NI with several choices for the present age of
the disk (10 \& 13 Gyr) and for the chemical composition of the infalling gas
(0.7 $\leq$ X$_{2inf}$/X$_{2P} \leq$ 1.0; 0.8 $\leq$ X$_{3inf}$/X$_{3P} \leq$
1.2).
As anticipated by RST, after the first few Gyr of evolution, stellar production
of  $^3$He dominates (see, e.g.,
Figure 8).  This contribution from newly synthesized $^3$He quickly overwhelms
the primordial contribution and, even in the absence of any primordial $^3$He
(and D), we predict solar system and ISM $^3$He abundances in excess of those
inferred from the observational data (see also Galli et al. 1995, Tosi et al.
1995).  Clearly, there is a problem and, until
this conflict is resolved, $^3$He cannot serve as a probe of BBN (RST).  The
problem may lie with the observational data and/or its interpretation.  Or, it
could be that our models - or some of the ingredients therein - are the
culprits.

  The solar system $^3$He data (meteorites, lunar soil and rocks, solar wind)
have recently been reanalyzed by Geiss (1993) and by Copi, Schramm \& Turner
(1995).  Although the size of the error estimates has increased compared, e.g.,
to those used in ST92, the central values remain unchanged.  Consistency  with
our model predictions would require that the presolar $^3$He abundance has been
underestimated by more than a factor of 1.5-2.0.

  The ISM data are more problematic.  If, indeed, stellar production of new
$^3$He is occurring as indicated by the planetary nebulae observations (Rood,
Bania \& Wilson 1992; RBWB95), then X$_{3}$ should be higher where
there is more stellar processing - in the inner galaxy.  As may be seen in
Figure 10, there is no observational evidence for such a trend.  Indeed, the
highest $^3$He abundances are derived from data for the HII regions in the
outer galaxy.  The possibility that the $^3$He abundances inferred from HII
region radio observations are not reliable indicators of the current ISM
values has also been invoked. Olive et al. (1995) have shown that
unresolved structures in the nebulae may lead to an
underestimate of their actual $^3$He content but, only by a few tenths.
On the other hand, the occurrence of a strong, $^3$He depleted, Wolf-Rayet wind
can reduce the $^3$He abundance inside the HII region, although we presume
that the external layers of many HII regions should still be uncontaminated and
that a negative internal gradient should appear in the abundances derived
from central to outer parts within the nebulae observed with sufficient spatial
resolution. Besides, if an HII region exhibits depleted $^3$He because of the
pollution from the central Wolf-Rayet star, it should also show enhanced
$^4$He, which does not seem to be the case for the RBWB95 sample. Olive et al.
(1995) have tried to understand the $^3$He
distribution as a reflection of the mass of the HII regions (more destruction
of $^3$He in the more massive HII regions).  This radical explanation makes
little sense to us (and, to them as well) since it would require that HII
regions are more efficient processors of interstellar material than the
galaxy as a whole (since more than half the gas in these HII regions would
have had to be cycled through the massive stars of the individual regions).
Since the HII region data are hard to acquire and difficult to analyze (see,
 e.g., RBWB95), work on both fronts is important and, beyond the
scope of our analysis here.  So, here we have adopted the extant data, assumed
that the inferred
abundances are correct and investigated the implications for our models.

  As may be seen in the Figures 8 \& 9, the contribution from newly synthesized
$^3$He is large compared to the differences among the various model assumptions
regarding the age of the disk, the infall abundances, the primordial abundances,
the specific models (1, 25, NI).  Therefore, the prime suspect must be our
estimates of the $^3$He yields from low and intermediate mass stars (Rood 1972
; Galli et al. 1994; Hogan 1995;
Wasserburg, Boothroyd \& Sackmann 1995).  To explore this avenue we have
considered a series of non-standard alternatives to our standard models.
Indeed, several suggestions have been published for physical mechanisms
to suppress the overproduction of stellar $^3$He.  They are related to
low-energy
resonances or to deep convective mixing (e.g. Galli et al. 1994; Hogan 1995;
Wasserburg, Boothroyd \& Sackmann 1995), but none of them seems  fully
consistent with all the available data.
In our cases II, III, VI and VIII the
suppression of stellar produced $^3$He may be sufficient to flatten the X$_{3}$
vs. t relation enough so that consistency  with the local data may be found
provided that the primordial abundances are small enough.  However,
it must be emphasized that if the detections of excess $^3$He in planetary
nebulae reported by Rood, Bania \& Wilson (1992) and RBWB95 are
confirmed, some of these alternatives (II and VI) are excluded.
By comparing the model predictions with all the available constraints, we
find that only case III is sufficiently compatible with all the data,
despite its shallow negative gradient in the X$_{3}$ vs. R relation.
In this case we may use
the solar system data to bound the primordial abundance of $^3$He from above
and, the nucleon-to-photon ratio from below.

 \vspace{1.5cm}
\centerline {7. CONCLUSIONS}

  We have tracked the evolution of the abundances of D and $^3$He in a variety
of chemical evolution models which incorporate the results of a newly computed
grid of stellar structure and evolution models.  We have confirmed that D is
fully destroyed during the pre-main sequence evolution and, therefore, its
galactic evolution is simple.  For the range of our models (see \S 3) we find
only modest destruction of D and, using the solar system and ISM data in
concert with our ``best" models, we bound the primordial (pre-galactic disk)
abundance,
        $$ 2.7 \leq 10^5 {\rm X}_{2P} \leq 6.6  ,\     1.8 \leq 10^5{\rm
y}_{2P} \leq
4.4.  \eqno (1)$$
For consistency with the predictions of standard BBN (WSSOK), we require that
the universal ratio of nucleons (baryons)-to-photons lie in the range,
     $$ 4.1 \leq \eta_{10} \leq 7.1.  \eqno (2)$$
In terms of the baryon density parameter, $\Omega_B$, $(\Omega_B$h$_{50}^2$ =
0.015$\eta_{10}$; H$_0$ = 50h$_{50}$ km/s/Mpc),
         $$ 0.06 \leq \Omega_B{\rm h}_{50}^2 \leq 0.10.  \eqno (3)$$
For our NI models the X$_{2P}$ estimate is somewhat higher and the bounds on the
baryon density slightly lower.

  Our stellar models, which account for the evolution of $^3$He (destruction,
survival, production) reveal that for low mass stars the production of newly
synthesized $^3$He is very important for the evolution of galactic $^3$He.
Indeed,
for all of our standard models (\S 4) production of new $^3$He is so dominant
that even in the absence of any pre-galactic D and/or $^3$He, $^3$He is
overproduced
compared to the solar system and ISM data.  Until this conflict is resolved it
is difficult to see how $^3$He can be used as a probe of BBN (RST).  Setting
aside
the possibility that the problem lies with the observational data, we have
explored a series of non-standard models (\S 5) in which some or all of the
newly
sysnthesized $^3$He is assumed to be destroyed before being returned to the ISM.
We find that if this suppression of $^3$He occurs only in stars $\leq$ 1
M$_{\odot}$,
 the depletion is ``too little, too late" to resolve the discrepancy with the
solar system data.  In contrast, $^3$He destruction in stars with masses in the
range 1-2
M$_{\odot}$ (Wasserburg, Boothroyd \& Sackmann 1995) can reconcile our models
with the data
provided that the initial $^3$He abundance is not too large.  In this case we
recover a lower bound to the baryon density which is consistent with that
derived from the D evolution data (see eqs. 1-3).  It remains
to be seen whether this ``fix" will be permitted by the PNe and
or $^{12}$C/$^{13}$C data.

\noindent
{\bf Acknowledgements}

\noindent
M.T. warmly thanks Letizia Stanghellini, Daniele Galli and Claudio Ritossa
and G.S. thanks M. Pinsonneault for  valuable and stimulating
discussions.  The work of G.S. is supported at Ohio State by the
Department of Energy (DE-AC02-76-ER01545); part of this work
was done while G.S. was a visitor at the Instituto Astronomico
e Geofisico (Sao Paulo, Brasil) and he thanks them for hospitality and
assistance.

\vfill\eject
\centerline {REFERENCES}
\par\noindent
Audouze, J. \& Tinsley, B.M. 1974, ApJ, 192, 487
\par\noindent
Balser, D.S., Bania, T.M., Brockway, C.J., Rood, R.T. \& Wilson, T.L. 1994,
 ApJ, 430, \par 667
\par\noindent
Bahcall, J.N.  \& Ulrich, R.K. 1988, Rev. Mod. Phys., 60, 297
\par\noindent
Bania, T.M., Rood, R.T. \& Wilson, T.L. 1987, ApJ, 323, 30
\par\noindent
Black, D.C.  1971, Nature Phys. Sci., 234, 148
\par\noindent
Black, D.C. 1972, Geochim. Cosmochim. Acta, 36, 347
\par\noindent
Bolte, M. 1994, ApJ, 431, 223
\par\noindent
Carswell, R.F., Rauch, M., Weymann, R.J., Cooke, A.J. \& Webb, J.K. 1994
 MNRAS, 268, \par L1
\par\noindent
Charbonnel, C. 1995, preprint
\par\noindent
Dearborn, D.S.P. 1992, Phys. Rep., 210, 367
\par\noindent
Dearborn, D.S.P. \& Eggleton, P.P. 1976, QJRAS, 17, 448
\par\noindent
Dearborn, D.S.P., Griest, K. \&  Raffelt, G. 1991, ApJ, 368, 626
\par\noindent
Dearborn, D.S.P., Schramm, D.N. \& Steigman, G. 1986, ApJ, 203, 35 (DSS)
\par\noindent
Edmunds, M.G. 1994, MNRAS, 270, L37
\par\noindent
Eggelton, P.P. 1967, MNRAS, 135, 243
\par\noindent
Eggelton, P.P. 1968, MNRAS, 143, 87
\par\noindent
Fields, B.D. 1995, ApJ, In Press
\par\noindent
Galli, D., Palla, F., Ferrini, F. \& Penco, U. 1995, ApJ, 443, 536
\par\noindent
Galli, D., Palla, F., Straniero, O. \& Ferrini, F. 1994, ApJ, 432, L101
\par\noindent
Geiss, J. 1993 in {\it Origin and Evolution of the Elements}, N.Prantzos,
 E.Vangioni-Flam \par and M.Cass\' e eds. (CUP, U.K.), p.89
\par\noindent
Geiss, J. \& Reeves, H. 1972, A\&A, 18, 126
\par\noindent
Giovagnoli, A. \& Tosi, M. 1995, MNRAS, 273, 499
\par\noindent
Hogan, C.J. 1995, ApJ, 441, L17
\par\noindent
Huebner, W.F., Merts, A.L., Magee, N.H.Jr., \& Argo, M.F. 1977, Astrophysical
\par Opacity Library (LASL, LA-6760-M)
\par\noindent
Iben, I.Jr. 1967, ApJ, 147, 624
\par\noindent
Iglesias, C. A. \& Rogers, F. J. 1990 , ApJ, 360, 221
\par\noindent
Iglesias, C. A. \& Rogers, F. J. 1992 , ApJ, 397, 717
\par\noindent
Linsky, J.L., Brown, A., Gayley, K., Diplas, A., Savage, B.D., Ayres, T.R.,
 Landsman, \par W., Shore, S.W. \& Heap, S. 1993, ApJ, 402, 694
\par\noindent
Matteucci, F. \& Fran{\c c}ois, P. 1989, MNRAS, 239, 885
\par\noindent
Olive, K.A., Rood, R.T., Schramm, D.N., Truran, J. \& Vangioni-Flam, E. 1995,
 \par ApJ, 444, 680
\par\noindent
Olive, K.A. \& Steigman, G. 1995, ApJS, 97, 49
\par\noindent
Pagel, B.E.J., Simonson, E.A., Terlevich, R.J. \& Edmunds, M.G. 1992, MNRAS,
 255, 325
\par\noindent
Prantzos, N. 1995, A\&A In Press
\par\noindent
Rood, R.T. 1972, ApJ, 177, 681
\par\noindent
Rood, R.T., Bania, T.M. \& Wilson, T.L. 1992, Nature, 355, 618
\par\noindent
Rood, R.T., Bania, T.M., Wilson, T.L.\& Balser, D.N. 1995 in {\it The Light
Element \par Abundances}, P.Crane ed. (Springer-Verlag), p. 201, RBWB95
\par\noindent
Rood, R.T., Steigman, G. \& Tinsley, B.M. 1976, ApJ, 207, L57, RST
\par\noindent
Scully, S.T. \& Olive, K.A. 1995, ApJ, In Press
\par\noindent
Skillman, E.D., Terlevich, R.J., Kennicutt, R.C., Garnett, D.R.
 \& Terlevich, E. 1994, ApJ,
 \par 431, 172
\par\noindent
Songaila, A., Cowie, L.L., Hogan, C.J., \& Rugers, M. 1994, Nature, 368, 599
\par\noindent
Spite, F. \& Spite, M. 1982, A\&A, 115, 357
\par\noindent
Steigman, G. 1994, MNRAS, 269, P53
\par\noindent
Steigman, G. \& Tosi, M., 1992, ApJ, 401, 150, ST92
\par\noindent
Steigman, G. \& Tosi, M., 1995, ApJ, 453, 173
\par\noindent
Thomas, D., Hata, N., Scherrer, R., Steigman, G. \& Walker,
T. 1996, In Preparation
\par\noindent
Thorburn, J. 1994, ApJ, 421, 318
\par\noindent
Tinsley, B.M. 1980, Fund.Cosmic Phys., 5, 287
\par\noindent
Tosi, M., 1988a, A\&A, 197, 33
\par\noindent
Tosi, M., 1988b, A\&A, 197, 47
\par\noindent
Tosi, M., Steigman, G. \& Dearborn, D.S.P. 1994, in {\it The Light Element
 Abundances}, \par P.Crane ed. (Springer-Verlag), p.228
\par\noindent
Tytler, D. 1995 in {\it QSO Absorption Lines}, G.Meylan ed. (Springer-Verlag)
In Press
\par\noindent
Vangioni-Flam, E. \& Audouze, J. 1988, A\&A, 193, 81
\par\noindent
Vangioni-Flam, E., Olive, K.A. \& Prantzos, N. 1994, ApJ,  427, 618
\par\noindent
Vassiliadis, E. \& Wood, P.R. 1993, ApJ, 413, 641
\par\noindent
Walker, T.P., Steigman, G., Schramm, D.N., Olive, K.A. \& Kang, H.S. 1991, ApJ,
 \par 376, 51,  WSSOK
\par\noindent
Wannier, P.G. 1980, ARA\&A, 18, 399
\par\noindent
Wasserburg, G., Boothroyd, A. \& Sackmann, I.-J., 1995, ApJ,  447, L37
\par\noindent
Yang, J., Turner, M.S., Steigman, G., Schramm, D.N. \& Olive, K.A. 1984, ApJ,
 \par 281, 493,  YTSSO

\vfil\eject

\centerline {FIGURE CAPTIONS}

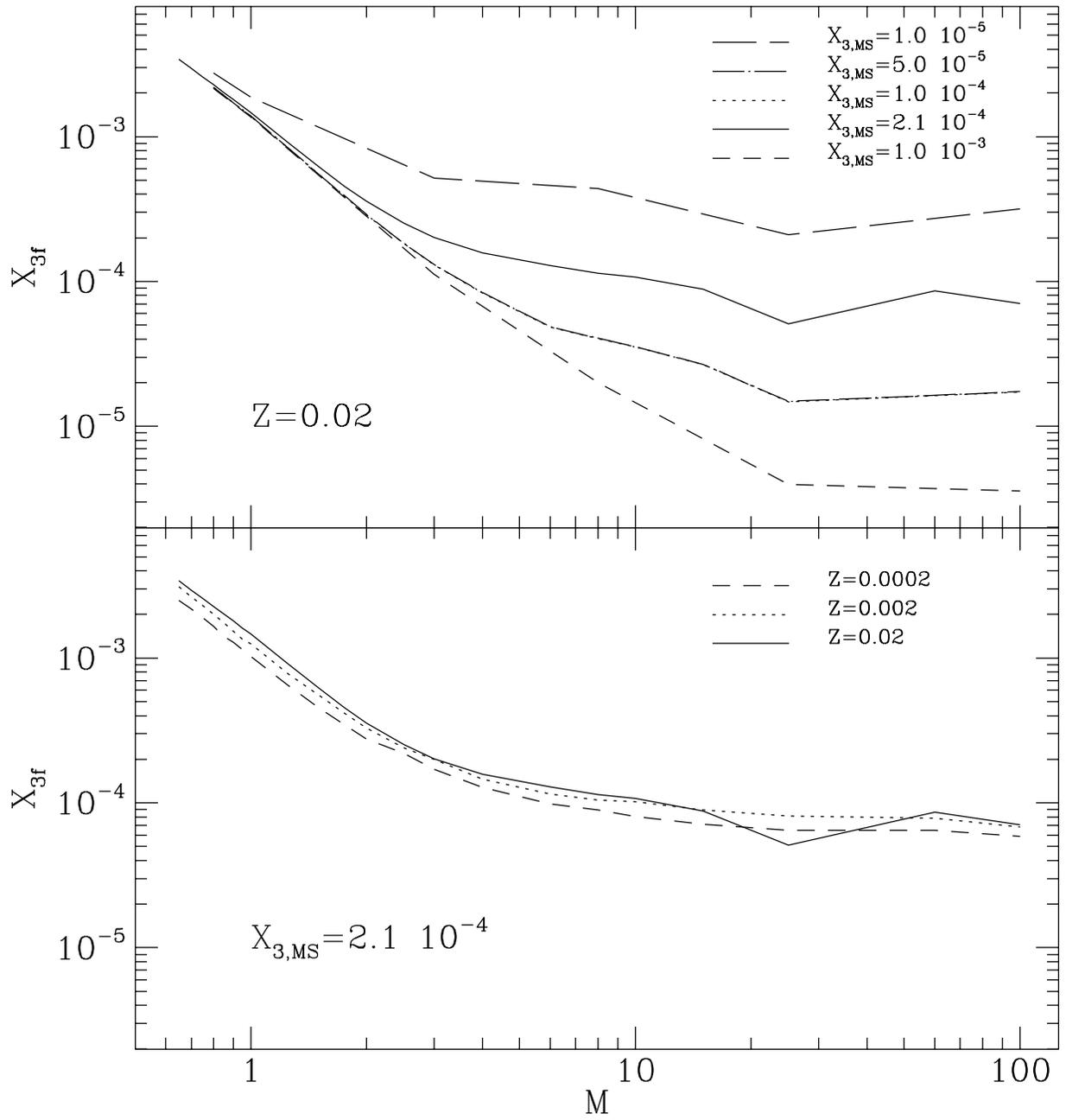
Figure 1. Top panel: final envelope abundance of $^3$He as a function of the
stellar initial mass for five different main sequence abundances of $^3$He
and Z=0.02. Bottom panel: final envelope abundance of $^3$He as a function of
the stellar initial mass for three different metallicities and $X_{3,MS}=2.1
\times 10^{-4}$.

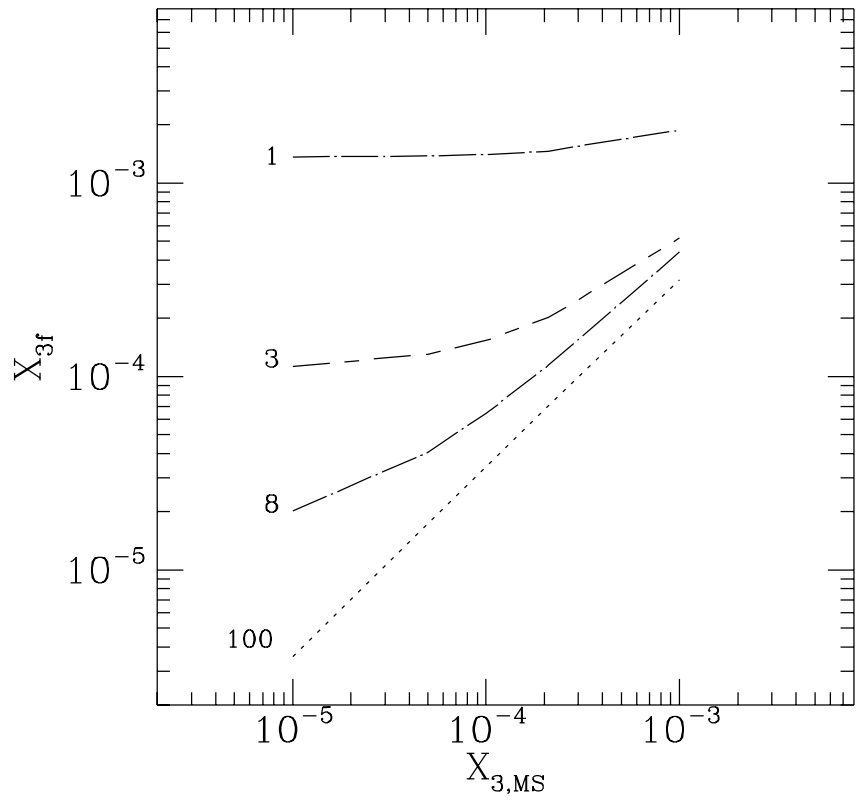
Figure 2. Final envelope abundance of $^3$He for Z=0.02 in stars of initial
mass $M/M_{\odot}$=1, 3, 8 and 100 as a function of the
main sequence abundance.

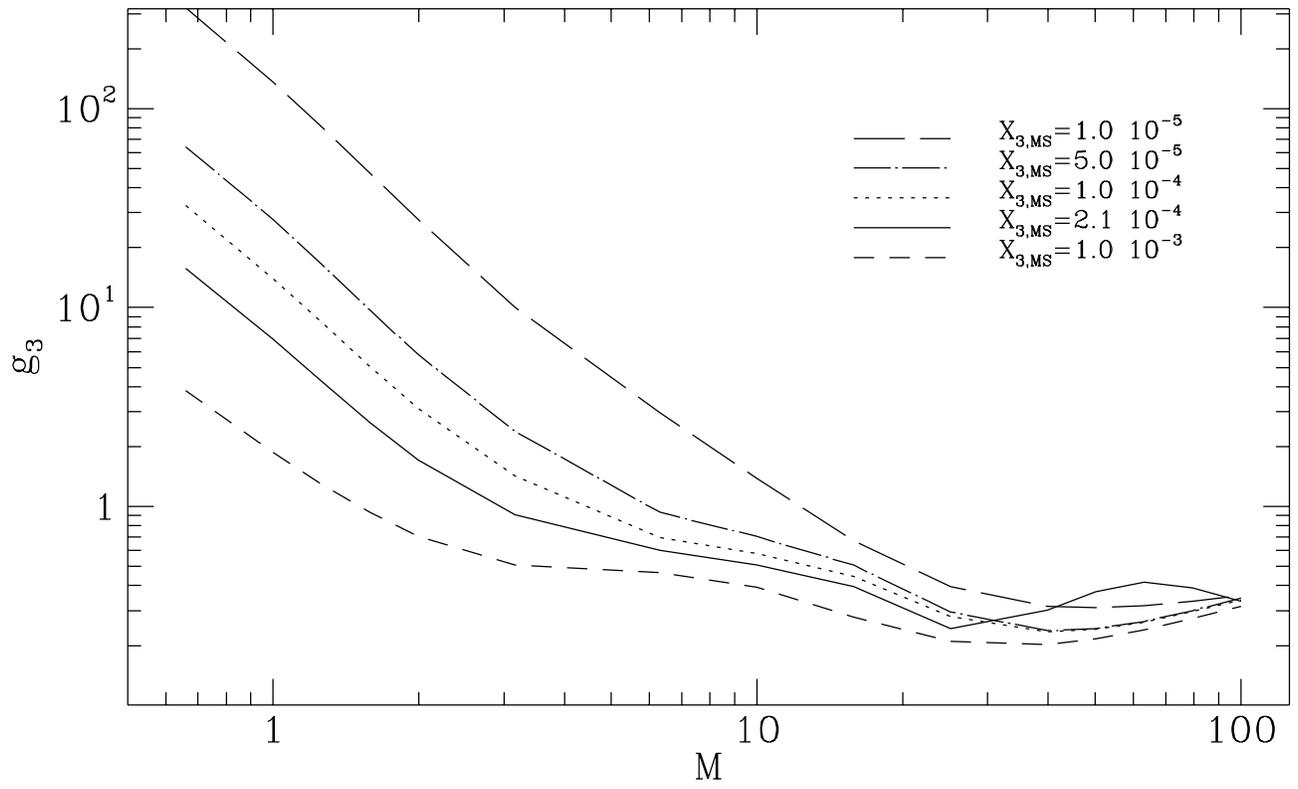
Figure 3. Survival fraction $g_3$ of $^3$He as a function of the
stellar initial mass for Z=0.02 and five different main sequence abundances of
$^3$He.

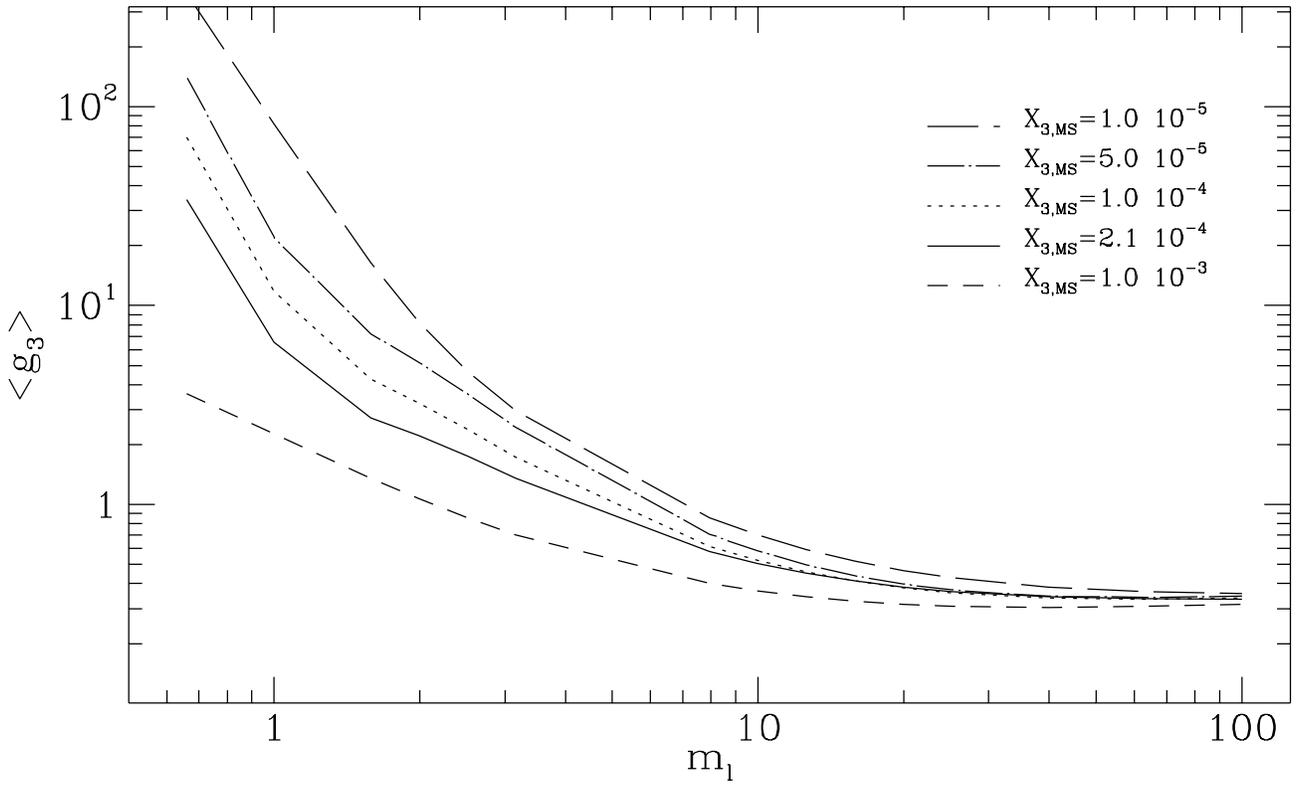
Figure 4. Average $g_3$ convolved with Tinsley's (1980) IMF as a function of
the lower mass limit $m_l$ on the IMF integral.

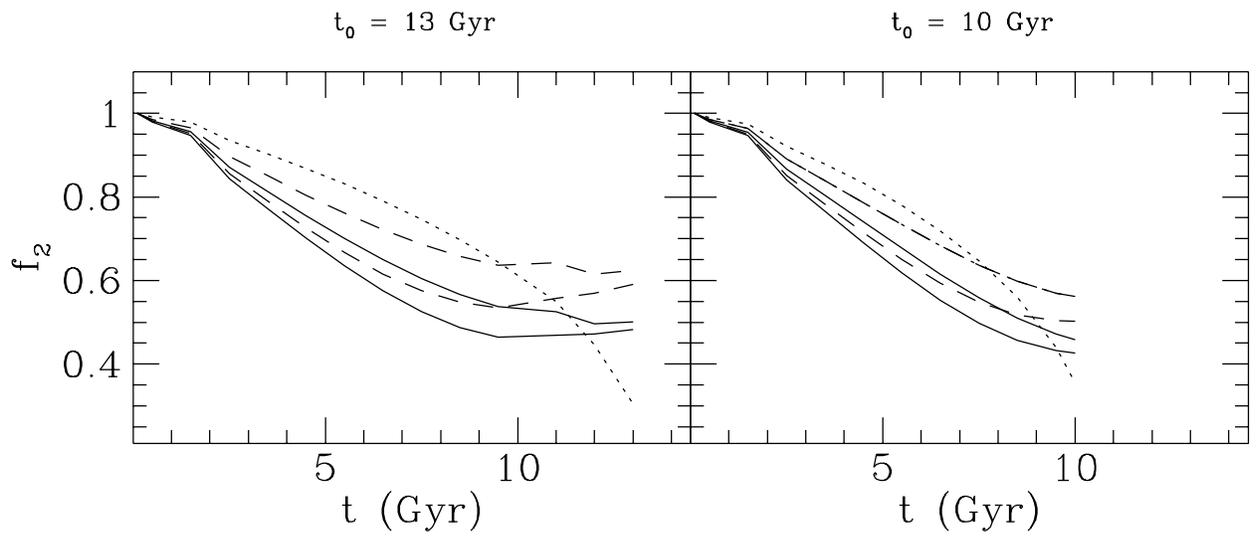
Figure 5. Deuterium survival factor $f_2$ as a function of time
predicted by different models of the solar ring (R= 8 kpc). Models shown in the
left panel assume 13 Gyr for the current age of the galactic disk; those
in the right panel assume 10 Gyr. In both panels, dotted curves correspond
to no infall models, solid curves to
models 1 (upper curve) and 25 (lower curve) with
$X_{2inf}=0.8X_{2P}$, dashed curves to models 1 (upper curve) and 25 (lower
curve) with $X_{2inf}=X_{2P}$.

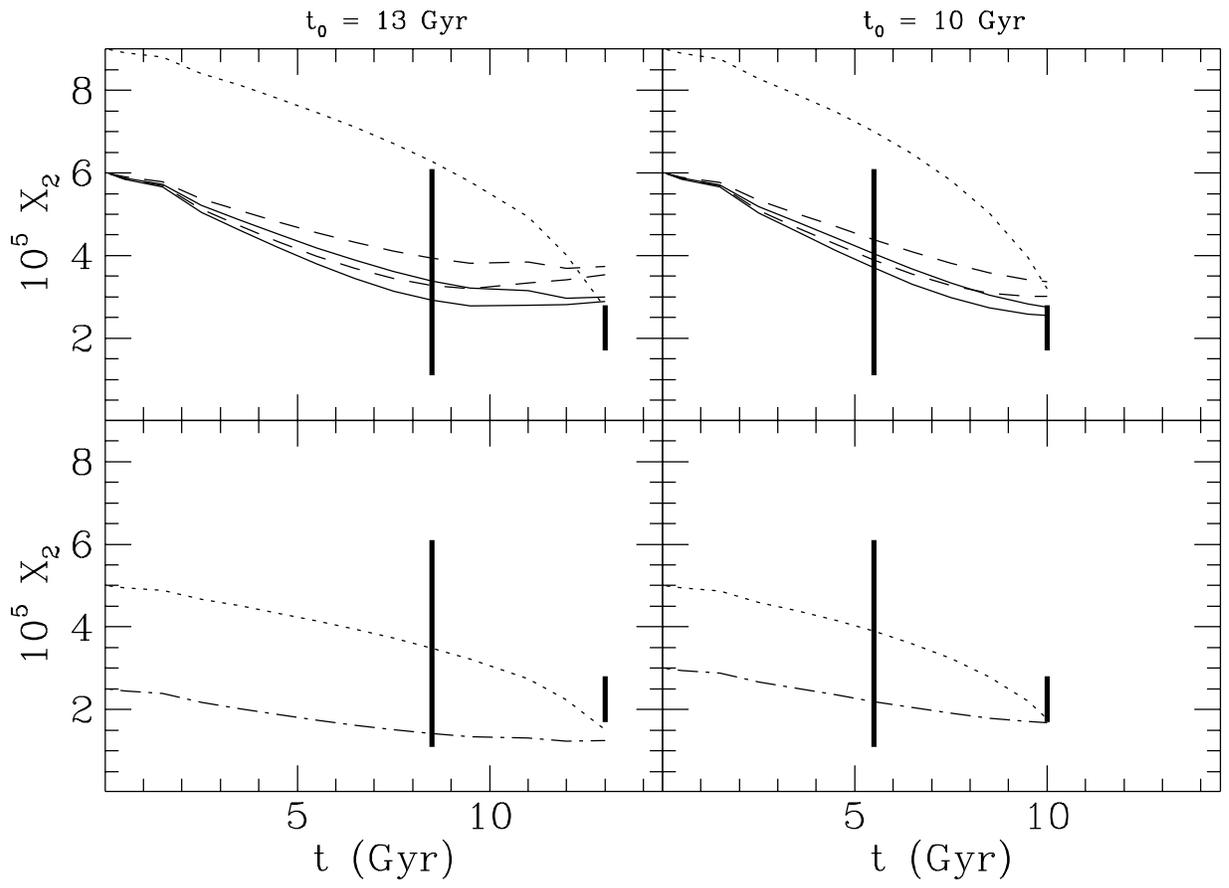
Figure 6. Evolution in the solar ring of the D abundance. Vertical bars give
the 2$\sigma$ range for the abundances derived from solar system and local
ISM observations (see text). Left panels
correspond to a disk age of 13 Gyr, right panels to 10 Gyr. Models in the
top panels assume the maximum value of primordial D consistent with the
data; models in the bottom panels the corresponding minimum value. The
dash-dotted lines correspond to ``standard" models with primordial infall
(1-B-Ia and 1$_{10}$-B-Ia), the other symbols are as in Fig.5

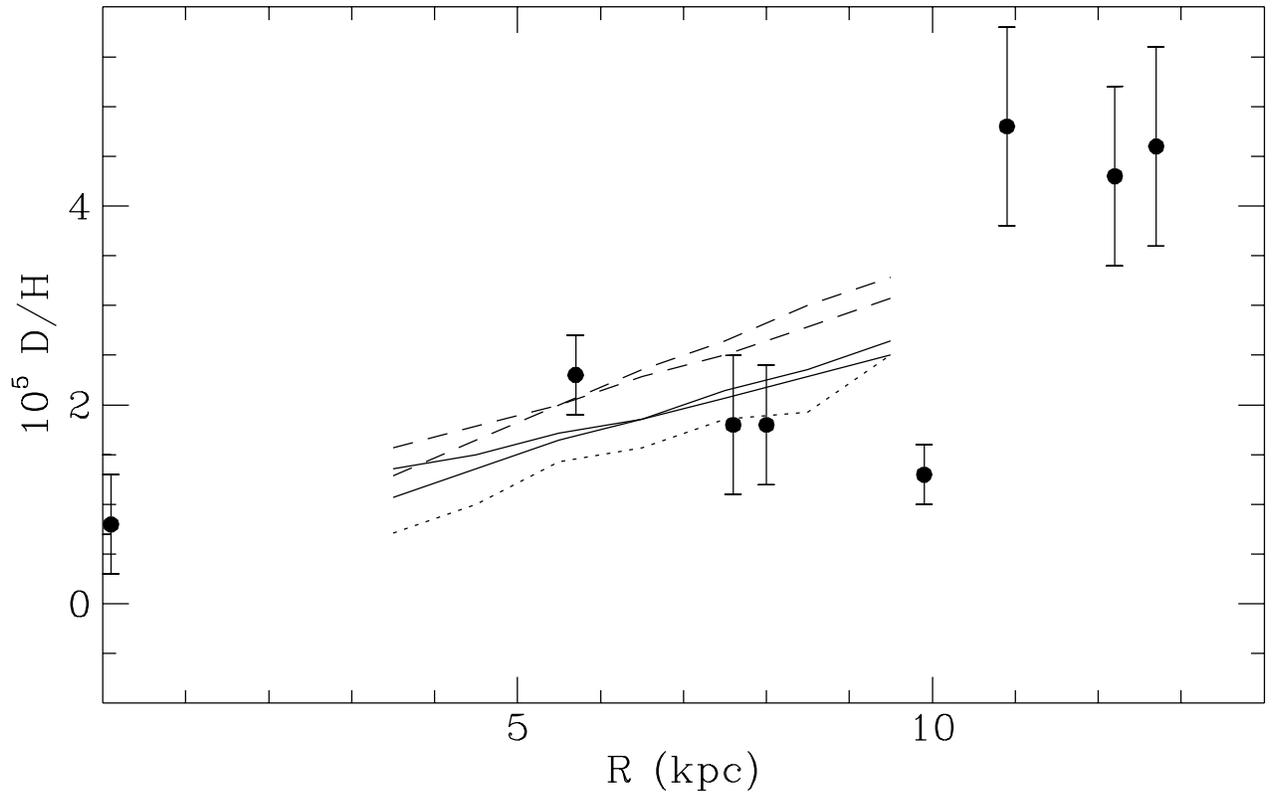
Figure 7. Current radial distribution of the D abundance resulting from the
same models as Fig.5 (same symbols as in that figure). In each pair of
curves of the same line-type, the steeper one corresponds to model 1 and the
flatter one to model 25. Data points are from Wannier (1980) but arbitrarily
divided by 100.

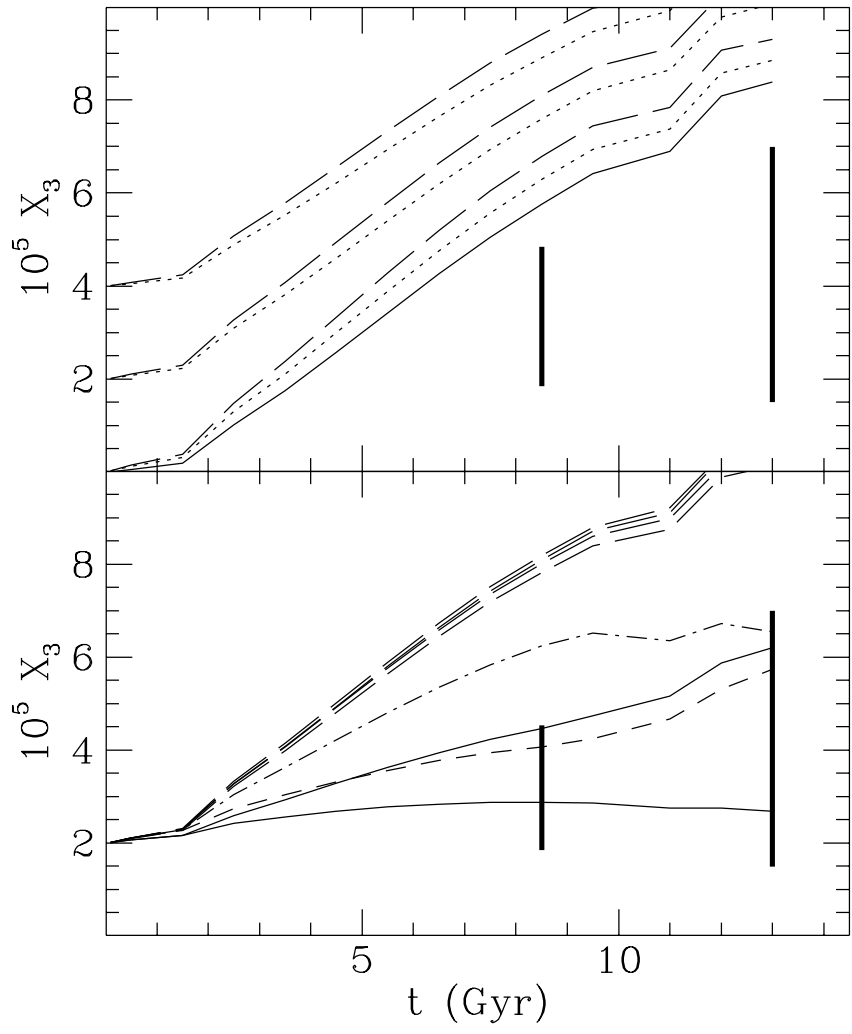
Figure 8.  Evolution of the $^3$He abundance in the solar ring as predicted
           by models 1.  The vertical bars show the 2$\sigma$ range for the
           solar system and ISM abundances (see text for details).  In the
           top panel all models adopt the standard stellar yields (see Sec.2
           and Table 1) and assume primordial infall.  They are in order of
           decreasing $^3$He and D abundances (i.e. from top to
           bottom): 1-R-Ia, 1-J-Ia, 1-W-Ia, 1-I-Ia, 1-X-Ia, 1-H-Ia, 1-M-Ia. 
	   The models in the lower panel
           all start with the same primordial abundances of D and $^3$He but
           have different adopted stellar yields and infall metallicity.  From
           top to bottom they are: 1-W-(Ic, Ia, Ib, Id, Vb, VIIIb, IIIb, VIb).
           
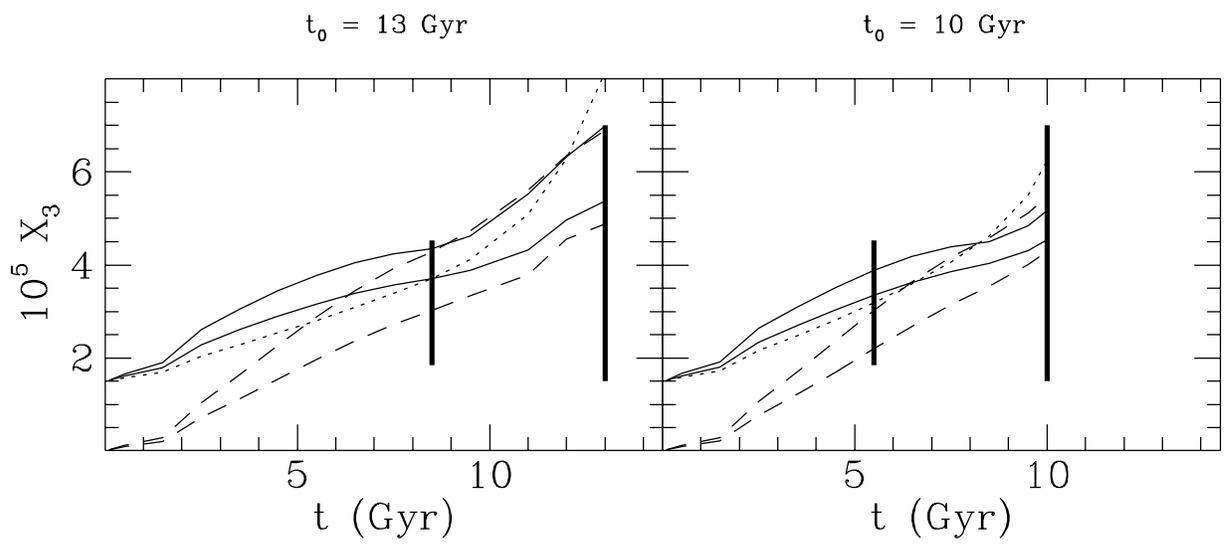
Figure 9.  Evolution of the $^3$He abundance in the solar ring as predicted by
           different chemical evolution models (1, 25, NI) using the same
           stellar yields and infall metallicity.  The left-hand panel
           corresponds to a disk age of 13 Gyr and the right-hand panel to 10
           Gyr.  The lower and upper solid lines are models 1-T-IIIb and 
           25-T-IIIb respectively (see Table 1); the lower and upper dashed 
           lines are models 1-H-VIIIb and 25-H-VIIb; the dotted line is for 
           NI-Z-III.

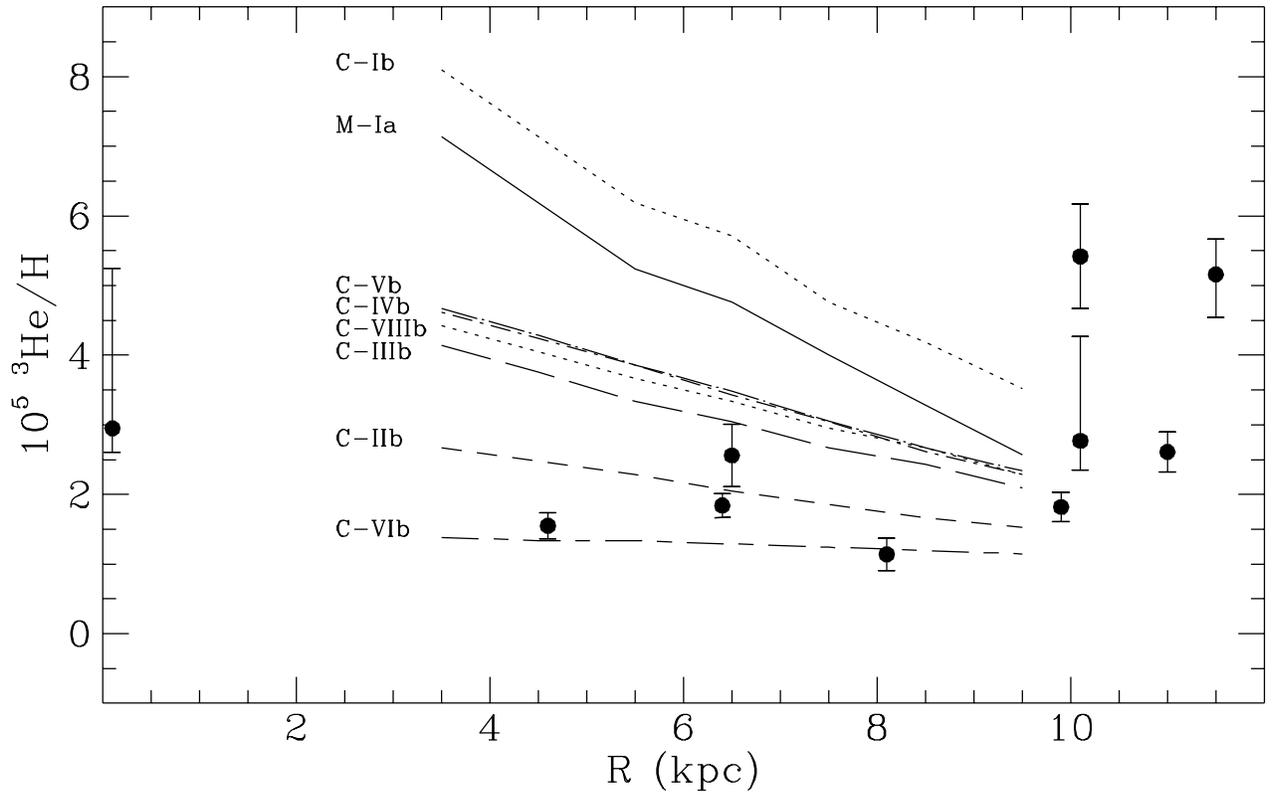
Figure 10. Predicted current radial distribution of the $^3$He/H number ratio.
Data points and error bars are from the HII region analysis by RBWB95 and the
specific models are labelled.

\vfil\eject
\voffset -1truecm
\renewcommand{\baselinestretch}{1.0}
\footnotesize
\thispagestyle{empty}
\vsize 28truecm
\hsize 15truecm
\begin{center}
\centerline {{\bf Table 1.} Models and Initial Abundances.}
\tabskip=1em plus.5em minus.5em
\begin{tabular}{lccclcc}
\hline
\\
Model$^1$ &10$^5$X$_{2p}$ &10$^5$~X$_{3p}$ & &
Model$^1$ &10$^5$X$_{2p}$ &10$^5$X$_{3p}$  \\
\\
\hline
\\
\multicolumn{7}{c}{Models with standard $^3$He evolution and primordial 
infall} \\
\\
1-M-Ia~~~~~~ & 0.0~~~~ & 0.0 &&   1-K-Ia~~     & 4.0~~~ & 1.0 \\
1-A-Ia~~     & 2.5~~~~ & 2.0 &&   1-O-Ia~~     & 4.5~~~ & 0.0 \\ 
1-H-Ia~~     & 3.0~~~~ & 0.0 &&   1-C-Ia~~     & 5.0~~~ & 2.0 \\ 
1-I-Ia~~     & 3.0~~~~ & 2.0 &&   1-X-Ia~~     & 6.0~~~ & 0.0  \\
1-B-Ia~~     & 3.0~~~~ & 2.5 &&   1-W-Ia~~     & 6.0~~~ & 2.0  \\
1-J-Ia~~     & 3.0~~~~ & 4.0 &&   1-R-Ia~~     & 6.0~~~ & 4.0    \\
1-L-Ia~~     & 3.85~~~ & 0.0 &&   1$_{10}$-B-Ia~~~~ & 3.0~~~ & 2.5 \\
\\
\multicolumn{7}{c}{Models with standard $^3$He evolution and 
non-primordial infall}  \\
\\
1-K-Ib~~     & 4.0~~~~ & 1.0 &&   1-W-Ib~~       & 6.0~~~ & 2.0    \\
1-K-Ic~~     & 4.0~~~~ & 1.0 &&   1-W-Ic~~       & 6.0~~~ & 2.0   \\
1-K-Id~~~~~~ & 4.0~~~~ & 1.0 &&   1-W-Id~~~~~~   & 6.0~~~ & 2.0   \\
1-C-Ib~~     & 5.0~~~~ & 2.0 &&   1-W-Ie~~       & 6.0~~~ & 2.0  \\
\\
\multicolumn{7}{c}{Models with non-standard $^3$He evolution and primordial infall} \\
\\
1-T-IIIa        & 6.0~~~~ & 1.5 && 25-T-IIIa       & 6.0~~~    & 1.5  \\
1$_{10}$-T-IIIa~~ & 6.0~~~~ & 1.5 &&  25$_{10}$-T-IIIa~& 6.0~~~    & 1.5  \\
\\
\multicolumn{7}{c}{Models with non-standard $^3$He evolution and 
non-primordial infall} \\
\\
1-H-VIIIb  & 3.0~~~~ & 0.0  &&   1-Q-IIIb   & 6.0~~~   & 3.0 \\
1-K-Vd     & 4.0~~~~ & 1.0  && 	1-Q-Vb   & 6.0~~~ & 3.0 \\
1-K-VIIIb  & 4.0~~~~  & 1.0 &&  	1-U-VIb    & 6.0~~~  & 3.5 \\
1-V-Vb     & 5.0~~~~  & 0.0 &&   1-R-VIb    & 6.0~~~  & 4.0 \\
1-S-Vb     & 5.0~~~~  & 1.0 &&   1$_{10}$-H-VIIIb& 3.0~~~   & 0.0\\
1-C-IIb    & 5.0~~~~  & 2.0 &&   1$_{10}$-C-VIb  & 5.0~~~   & 2.0 \\
1-C-IIIb   & 5.0~~~~  & 2.0 && 	1$_{10}$-T-IIIb & 6.0~~~    & 1.5 \\
1-C-IVb    & 5.0~~~~  & 2.0 && 	1$_{10}$-W-VIIIb & 6.0~~~    & 2.0 \\
1-C-Vb     & 5.0~~~~  & 2.0 &&   25-H-VIIIb        & 3.0~~~    & 0.0  \\	
1-C-VIb    & 5.0~~~~  & 2.0 &&   25-K-VIIIb        & 4.0~~~    & 1.0  \\
1-C-VII    & 5.0~~~~  & 2.0 &&   25-C-VIb        & 5.0~~~    & 2.0  \\
1-C-VIIIb  & 5.0~~~~  & 2.0 &&   25-C-VIIIb        & 5.0~~~    & 2.0  \\
1-T-IIIb   & 6.0~~~~  & 1.5 &&   25-T-IIIb       & 6.0~~~    & 1.5  \\
1-T-Vb     & 6.0~~~~  & 1.5 &&   25$_{10}$-H-VIIIb& 3.0~~~    & 0.0 \\
1-W-IIb  & 6.0~~~~   &2.0   &&   25$_{10}$-T-IIIb& 6.0~~~    & 1.5 \\
1-W-IIIb & 6.0~~~~   &2.0   &&   NI-C-III       & 5.0~~~    & 2.0 \\
1-W-Vb   & 6.0~~~~   &2.0   &&   NI-Z-III       & 9.0~~~    & 1.5 \\
1-W-VIb  & 6.0~~~~   & 2.0  && 	NI$_{10}$-C-III & 5.0~~~    & 2.0 \\
1-W-VIIIb~~~& 6.0~~~~ & 2.0 &&	NI$_{10}$-Z-III & 9.0~~~    & 1.5 \\
\\
\hline
\end{tabular}
\end{center}

$^1$Arabic numbers refer to the type of chemical evolution model (see text and
Tosi 1988a). Capital letters indicate the adopted primordial abundances.
Roman numerals indicate the stellar yields adopted for $^3$He: I, as 
described in section 2; II, as for Model I but with g$_3$=1 for 
M $<$ 2 M$_{\odot}$; III, as for Model I but with g$_3$=1 for stars with 
1$<$M/M$_{\odot}<$2; IV, as for Model I but with total $^3$He destruction
in stars with M $\leq$ 1 M$_{\odot}$; V, following Pinsonneault (1995); VI, 
following Hogan (1995); VII, as for Model VI but with $^3$He destruction for
M $\leq$ 1.6 M$_{\odot}$; VIII, as for Model VI but with $^3$He destruction in 
the range 1.3 $<$M/M$_{\odot}<$2.5. 
 Lower case letters indicate the adopted infall 
abundances: a,  X$_{2inf}$=X$_{2p}$ and X$_{3inf}$=X$_{3p}$; b, 
X$_{2inf}$=0.8X$_{2p}$  and X$_{3inf}$=X$_{3p}$; c, X$_{2inf}$=0.8X$_{2p}$ and 
X$_{3inf}$=1.2X$_{3p}$; d, X$_{2inf}$=0.8X$_{2p}$ and X$_{3inf}$=0.8X$_{3p}$;
e, X$_{2inf}$=0.7X$_{2p}$ and X$_{3inf}$=X$_{3p}$.

\end{document}